**A negative-*U* interpretation of the femto-second laser pulse induced crystallographic expansion of a cuprate HTSC material reported recently by Gedik *et al.***

**John A Wilson**

H.H. Wills Physics Laboratory

University of Bristol

Bristol BS8 1TL   U.K.

**Abstract**

Gedik *et al* have very recently demonstrated using a pump/probe femto-second laser technique that the *c*-axis lattice parameter of $LaCuO_{4+\delta}$ temporarily becomes expanded by as much as 2½% following pulsed laser optical excitation at 1.55 eV. Access to an out-of-equilibrium metastable excited state is observed to develop on a time scale of 30 ps. Subsequently the latter state decays displaying a still longer half-life of just over 300 ps. Observation of the temperature independence of this laser induced interstate transfer and of the linear dependence of the production of the metastable population upon the energy delivered per unit area by the initiating light pulse (beyond a key threshold fluence) have been interpreted by Gedik *et al* within the framework of standard p-to-d, O-to-Cu, 'charge transfer' excitations. By contrast these same data are reinterpreted here in terms of pumped local pairs, within a negative-*U* scenario of cuprate HTSC behaviour long advocated by the current author. The $d^8$-to-$d^{10}$ laser-induced augmentation in the negative-*U* state population ($^{10}Cu_{III}^{2-}$) brings marked *c*-axis expansion by virtue of (i) the local electrostatic charge imbalance, (ii) the increased antibonding nature of the electron double-loading $d^{10}(p^6)$ configuration created at pair-receptive $Cu_{III}$ coordination units, and (iii) the layered nature of the cuprate crystal structure. The new observations are related through to Röhler's striking, standard crystallographic observations, to stripe domain formation, and to previous pump/probe experiments.

**Key words**: mixed-valent cuprates, high temperature superconducting mechanism, negative-*U* states, pump/probe techniques.

### §1. Summary of the pump/probe experiment of Gedik *et al*.

In an interesting alternative to customary pump/probe, femto-second, pulsed laser spectroscopy, Gedik *et al* [1] have recently secured very revealing new information on the ultra-fast, local electronic/structural conditions in an excited high-$T_c$ cuprate superconductor. The material investigated was optimally-doped $La_2CuO_{4+\delta}$, in thin film form, and the probe technique used was electron diffraction of the excited state structure (as against the earlier examination of changes to the optical properties). The electron pulses employed to obtain this time-dependent diffraction information were generated by converting energy-tripled photoelectrons, produced under the initial 100 femto-second laser pulse, into a short wavelength electron packet via acceleration of these photoelectrons through a 30 kV potential drop. A de Broglie wavelength of 0.07 Å ensues, and this permits high-order small-angle electron diffraction data then to be acquired (in reflection) from the excited sample. Note the penetration depth for the diffracting electrons (at just 10 nm) is much less than that of the laser light (~ 150 nm). The $La_2CuO_{4+\delta}$ film (of thickness **XXX** nm) stands on a $LaAlSrO_4$ substrate, the latter transparent to the given laser radiation. The radiation is delivered at rather high fluences of up to 20 mJ cm$^{-2}$, but only at such a pulse repetition time (1 ms) as to ensure relatively little sample heating. The path difference travelled through the equipment between the light pulse and the slower electron packet is made such as to yield absolute relative time shifts between the two signal channels. The (0,0,10) *c*-axis spot locations were monitored in the equilibrium and out-of-equilibrium structural conditions of the sample. The diffraction patterns were recorded by use of a CCD camera sensitive to the receipt of individual electrons.

Very significant changes in the local *c*-axis lattice parameter have been discovered of as much as 2½%, far bigger than could arise simply from thermal expansion. The actual $c_o$ values attained within each cycle are dictated by the energy delivered to the sample per unit illuminated area by the change-initiating light pulse. The femto-second pulses come from a 1.55 eV photon energy, $Ti^{3+}$:sapphire laser. This energy is, it so happens, a special one for the cuprate HTSC materials, as made apparent in previous optical pump/probe spectroscopy [2],[3] and related experimentation [4],[5]. The present activating light train is of less than one pico-second duration. Below $T_c$ the bulk of the Cooper pairs reconstitute on a 10 ps time scale, but entry into the *c*-axis expanded, metastable condition proceeds over a somewhat longer characteristic settling-in period or 'rise time' with $\tau_{½} \approx 30$ ps. Subsequently this expanded condition decays manifesting a very much longer characteristic half-life of just over 300 ps. Accordingly the original condition is not fully restored until after several nano-seconds. Successive laser pulses were spaced out at intervals much longer than this (1 ms), and their intensity could be modified to probe the effect upon the observed structural response of the photon dosage to have been delivered per unit illuminated area (or 'fluence').

One very notable outcome is that the monitored transient diffraction changes are discovered to be virtually independent of temperature up to 300 K – a feature that speaks of their primarily electronically driven form. The lattice parameters locally become very appreciably changed (*c* by up to 2½%) without any alteration to the crystal structure or symmetry. There is observed to arise steady transfer of diffraction weighting across an isosbestic-like point (in $c^*$) between the coupled initial and expanded state spot locations ((0,0,10) normally being the spot followed) as the metastable product of the interstate transfer excitation accumulates. The ultimate fall-back in the latter population does not exhibit such a characteristic, but would seem from its temperature independent form to involve the emission of acoustic phonons. The clear isosbestic nature to the metastable state build-up permits quantitative record of the process to be obtained through an integration of the diffraction weighting changes occurring to either side of the isosbestic point, and then producing time-difference diffraction profiles. The actual magnitude of the maximum spot displacement attained depends upon the total photon dosage delivered per unit illuminated area to the sample per initiating light pulse. Maximal displacement, materializing around 30 ps after the initial pulse, is found to grow linearly with the fluence employed, but this rise begins only beyond a set threshold level of 5 mJ cm$^{-2}$. Very interestingly the latter (rather high) threshold fluence equates (upon insertion of an absorption coefficient of 7x10$^{-4}$ cm$^{-1}$) to the delivery per pulse of 0.12 of a 1.55 eV photon per planar Cu atom. At this juncture up to one in eight of the basal Cu-O coordination units could have absorbed such a photon, a figure, note, which is very comparable to the 'charge doping' level in the measured optimally-doped sample of La$_2$CuO$_{4+\delta}$. We shall return to this important observation in §4. The continuous changes detected in overall lattice parameter with the time lapsed and with the fluence delivered bring to mind isostructural alloying, and what there is known as Végard's law behaviour, as a sample strains cell by cell to accommodate to the local atomic size distribution.

## §2. Summary of the negative-*U* interpretation of the HTSC phenomenon.

In the publication by Gedik *et al* [1] the above metastable condition is proposed to come from standard p-to-d 'charge transfer' electronic excitation, but quite why the latter should in that case precipitate a metastable state of such unusually long duration is not apparent to the author. Furthermore, both optical data and band structural calculation would indicate that this normal charge transfer edge (from the topmost valence band states of dominantly O(p$_z$) to the lowest conduction band states of dominantly Cu(d$_{x2-y2}$) parentage) is not reached in the cuprates until 1.8 eV or above. What *is* encountered at somewhat lower energy in the mixed-valent cuprates are di- to tri-valent inter-coordination unit charge transfers. It is, of course, electron transfers like these that confer upon the HTSC systems, essentially mixtures of the two Mott insulators La$_2$CuO$_4$ and LaCuO$_3$, their mixed-valent metallicity. At a local level such inter-subsystem transfers (hops from coordination unit type A to B) can be expressed in the notation I introduced in [6] as

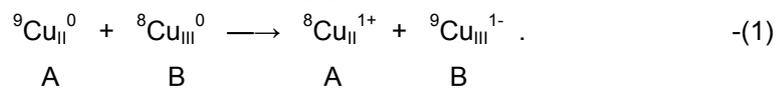
$$^9Cu_{II}^0 \; + \; ^8Cu_{III}^0 \; \longrightarrow \; ^8Cu_{II}^{1+} \; + \; ^9Cu_{III}^{1-} \; . \qquad -(1)$$
$$\quad A \qquad\quad B \qquad\qquad A \qquad\quad B$$

Here the Roman subscripts denote the formal, Cu(d)/O(p)-hybridized site valence, whilst the upper left numbers indicate the net complement of outer (d-symmetry) valence band electrons instantaneously residing in the given Cu-O coordination unit. The upper right figure marks then the site charge balance (0) or imbalance (for the higher valent $Cu_{III}$ site above, the 1- of an extra electron – the 'single loading' of the $d_{x2-y2}$ symmetry $\sigma^*$-state there).

A site partaking of formal trivalence is conferred within the mixed-valent milieu by proximity to some substituent or interstitial atom. Such augmentation in Cu valency is called for to satisfy locally the requisite electron complement still demanded by the main oxygen-dominated valence-band states (as in Sr-substituted $(La/Sr)_2CuO_4$), or to satisfy some increased requirement (as in interstitially super-oxygenated $La_2CuO_{4+\delta}$). Optimal doping in regard to HTSC $T_c$ value finds 16% of the CuO array of coordination units at any one time so raised in formal valency to $Cu_{III}$. This local nature to the doping is very evident when using local probes such as nmr.

The above local, inter-subsystem, electron transfer excitation was revealed from the earlier spectroscopic pump-probe [2-4] and related experiments [5] to occur in sharp excitonic fashion at – by chance – virtually 1.55 eV. Moreover this earlier pump-probe work first made manifest the ensuing population of relatively long-lived states being accessed beyond the initially excited condition. What was advocated in [6] by the present author as happening is entry into metastable 'double-loading' at the $Cu_{III}$ sites. One should for the above excited trivalent (B) sites consider an energetically down-hill' two-site relaxation as being open to the hot carriers

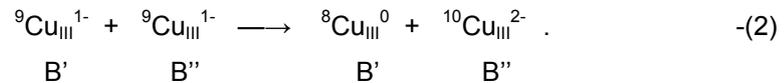
$$^9Cu_{III}^{1-} + {}^9Cu_{III}^{1-} \longrightarrow {}^8Cu_{III}^{0} + {}^{10}Cu_{III}^{2-} \quad . \qquad -(2)$$
$$\text{B'} \qquad \text{B''} \qquad \text{B'} \qquad \text{B''}$$

Such a mode of attainment of the doubly-loaded 'negative-$U$' state bears echoes of the permanent site 'disproportionation' exhibited by *divalent* AgO, or AuO and $CsAuCl_3$, where

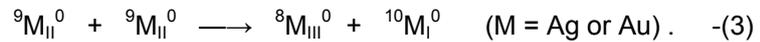
$$^9M_{II}^{0} + {}^9M_{II}^{0} \longrightarrow {}^8M_{III}^{0} + {}^{10}M_{I}^{0} \quad (M = Ag \text{ or } Au) \quad . \quad -(3)$$

In the latter process the crystal structure actually becomes statically transformed to set up formal $M_I$ and $M_{III}$ sites, these displaying changed local structural coordinations and associated Madelung energies. In the cuprates no such permanent disproportionation is directly accessible from the Mott-insulating *divalent* condition. Note that whether the disproportionation process (3) goes forward or not is dictated mainly by the degree of stability (in relation to $d^9$) being introduced with the $(p^6)d^{10}$ closed-shell configuration, this aided somewhat by the relative stability gained at $d^8$ upon adoption of four-fold square-planar coordination. What is different between the above referred to 4d/5d Ag/Au materials and the 3d cuprates is a diminished anti-bonding elevation for the $pd\sigma^*$ states within 'first series' materials, the result of their relatively contracted d-symmetry wavefunctions. Recall in a 3d material that there are present no 2d states within the atom core with which to necessitate state orthogonality. The outcome is a disparately contracted 3d-state radial expectation value – a condition bringing to first-series materials their increased propensity towards Mott localization. (A comparable situation arises later on in the Periodic Table between 4f and 5f systems). This 3d state binding contraction becomes very rapidly more pronounced, both radially and energetically, the greater the formal site valence in play. One immediate outcome with regard to the latter point is that ($^9$Zn$_{III}^0$) LaZnO$_3$ becomes totally inaccessible, with even ($^8$Cu$_{III}^0$) LaCuO$_3$ barely stable. Indeed it is because $^8$Cu$_{III}^0$, $^9$Zn$_{III}^0$, and beyond them $^{10}$Ga$_{III}^0$ have become so deep-lying – the latter in Ga$_2$O$_3$ and GaN falling at -18 eV – that a metastable, negative-$U$, 'double-loading' state $^{10}$Cu$_{III}^{2-}$ is ever to be attainable, bearing in mind the large Madelung potential change incurred due to its now reduced *nuclear* charge.

Process (2) discussed above would be one for producing local pairs from the optically generated excess content of $^9$Cu$_{III}^{1-}$. What about obtaining such pairs directly from $^9$Cu$_{II}^0$, the majority species in the mixed-valent, unpumped state, towards the sourcing at reduced temperatures of the system-wide superconductivity? What long ago was proposed in [6a,b] is the three-site, mixed-valence process

$$^9Cu_{II}^0 + {}^9Cu_{II}^0 + {}^8Cu_{III}^0 \longrightarrow {}^8Cu_{II}^{1+} + {}^8Cu_{II}^{1+} + {}^{10}Cu_{III}^{2-}. \quad\quad \text{-(4)}$$

This achieves the desired product, $^{10}$Cu$_{III}^{2-}$, (and through the agency of the latter the nodal Cooper pairs) without the energetically forbidden and structurally undesirable divalent disproportionation of process (3). By virtue of the crystal structure the greater part of the carriers in HTSC materials occupy states in the axial, saddle regions of the Fermi surface. These not only have their Fermi momenta directed largely in the *x* and *y* directions, but, in consequence of the saddle geometry, their Fermi velocities too are axially directed (*y* and *x*). Residing at the shallow saddles (for the given *n* and $E_F$) these carriers are heavy and highly correlated. Indeed they become prone to weak localization as *n* slips back towards unity (*p* → 0) and the Mott-insulating, band half-filling condition is approached. These are the carriers which undergo the chronic anisotropic electron-electron scattering that characterizes the general transport properties of the HTSC condition [6d]. The boson creation process of equation (4) is seen as sitting at the heart of this strong scattering together with the ensuing electron-boson scattering. In process (4) the dominantly *x*- and *y*-directed electrons are understood as entering onto the third, *trivalent*, site in a collision (inelastic wrt the lattice) cum joint tunnelling process (resonant wrt $E_F$) to provide the doubly-loaded $^{10}$Cu$_{III}^{2-}$ entities. The latter pair-state is, observe, not identical to $^{10}$Cu$_I^0$, as found in copper monovalent Cu$_2$O or KCuO – hence the usage of our expanded notation. Clearly $^{10}$Cu$_{III}^{2-}$ is to be expected to reside at a somewhat greater absolute binding energy than $^{10}$Cu$_I^0$ due to its local trivalent Madelung potential and ion radius. Indeed in figure 1 of [7] I positioned the $^{10}$Cu$_{III}^{2-}$ state to reside in close resonance with the Fermi energy, as the latter comes to be established within the mixed-valent Cu$_{II}$/Cu$_{III}$ oxide. This Feshbach-like resonant stationing of the 'negative-$U$' double-loading condition is analogous to what presently is being met with in the ultra-low temperature, gaseous, trapped atom/molecule systems, with their effective energy degeneracy between the fermionic and bosonic components to the overall system.

BCS/BEC, mixed system, crossover modelling of the HTSC phenomenon has long been addressed in certain theoretical works [8], but without ever examining too closely the sourcing of such a circumstance, or taking on board the mixed-valent, inhomogeneous nature of the situation. From a consideration of Stevens *et als'* optical pump/probe data [2], first in [7] and then subsequently in [9-14] (following the kinetic induction work of Li *et al* [4], the infra-red thermomodulation work of Little *et al* [5] and further pump/probe laser work from Demsar *et al* [3]), identification was afforded of the negative-*U* energy for the metastable $Cu_{III}$ based condition as being -1.5 eV per electron (or -3.0 eV per pair – as customarily this quantity is specified). As indicated, such a value would witness $^{10}Cu_{III}^{2-}$ being, by chance, stationed at the same absolute energy as $^{8}Cu_{III}^{0}$, the latter in unison with $^{9}Cu_{II}^{0}$ setting up the Fermi level in the mixed-valent metal. The overall situation is as summarized in figure 1 (slightly redrawn from [7] as in [9]).

The kinetic induction work of Li *et al* [4] proved to be especially valuable in fixing the above identification. This particular technique deals specifically with the superconductive behaviour and with the breakage of pairs. The laser which Li *et al* employed was tunable between 1.45 and 1.65 eV, and this flexibility permitted clear resolution of the excitonic nature of the main feature centred upon 1.54 eV to be obtained. Very significantly it enabled too the detection of a pair of satellite features, one lying to either side of the principal excitation and displaying very pronounced and distinctive temperature dependencies. The energies of the satellites are most revealing, standing in $YBCO_7$ at 41 meV to the low energy side and at 70 meV to the high energy side of the central peak. In [4] Li *et al* attempt to relate the above events again to 'charge transfer' excitation, but in [10–§3] I returned once more to negative-*U* based interpretation of these observations. There the 41 meV satellite was deemed associated with a local pair breaking in this case assisted by a spin-flip (singlet→triplet) energy contribution extracted from the reservoir of Cooper pair states during the negative-*U* pair disruption process

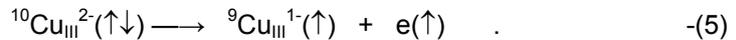
$$^{10}Cu_{III}^{2-}(\uparrow\downarrow) \longrightarrow \, ^{9}Cu_{III}^{1-}(\uparrow) \, + \, e(\uparrow) \quad . \quad \text{-(5)}$$

Appropriately this 41 meV satellite peak fades away as the temperature is elevated somewhat beyond $T_c$, mirroring here what is to be observed in neutron spin scattering – taken to track directly the singlet→triplet Cooper pair excitation [15]. The higher energy satellite is, by contrast, adjudged to be associated with the production of an optic phonon during the course of local pair disruption. The latter satellite becomes more pronounced this time as the temperature is raised. Its energy shift of 70 meV points to the Jahn-Teller extended Cu-O bond *c*-axis phonon as being the one here involved. As noted in [10] that particular phonon mode is strongly in evidence in I-R work in spite of the system being a metal. Why in particular this *c*-axis bond should become coupled with process (5) is that the excitation product $^{9}Cu_{III}^{1-}$ presents a Jahn-Teller active configuration.

### §3. The engagement of the lattice in HTSC events.

The above is very definitely not the only point at which one finds phonons – or, at a local level, the lattice – becoming enmeshed in HTSC phenomena. Indeed many phonons exhibit sharp changes at $T_c$, as manifest in I.R. and particularly Raman experiments [16]. Moreover techniques like EXAFS and PDF [17,18] directly register marked changes to the Cu-O bond lengths at $T_c$. Then again we encounter the now infamous 50-70 meV 'kinks' in the ARPES band dispersion data, in regard to which so much recently has been written [19,20,21]. In [11 & 13] I have discussed the latter extensively within the present negative-$U$ scenario, in terms of the local pair controlled condensate supporting at such an energy a plasmon mode of $A_{1g}$ symmetry. This mode is able then to couple with the topmost optic phonon, which this time relates to the *basal* Cu-O bonds. A strong dynamic coupling of the above local-pair-governed mode with the $\Sigma_1$ branch from this $B_{2g}$ phonon is very much in evidence in the neutron scattering derived, phonon dispersion curves towards the zone boundary [22]. Perhaps still more revealing, and now highly relevant to our understanding of the findings obtained by Gedik *et al* [1], are the direct and very accurate lattice parameter determinations put together quite recently by Röhler [23] right across several entire series of systematically doped HTSC systems.

What the Röhler data plots make manifest is that within the central range of HTSC-inducing dopings (*i.e.* from $p$ = 0.10 through to 0.22) the *basal* lattice parameters are significantly enlarged as compared with what one should expect were the measurements simply to follow a smooth decrease from $^9Cu_{II}^0$ to $^8Cu_{III}^0$, attributable to the increase in valence and the diminished number of antibonding electrons. The detected enlargement is observed to be greatest at $p$ = 0.16, precisely where $T_c$ maximizes, and it is, furthermore, larger for those systems for which $T_c^{max}$ is higher. The magnitude of this finding (recorded, it must be noted, up at 300 K) is around ½% in $a_o$. This should be compared with the enlargements of up to 2½% in $c_o$ witnessed now by Gedik *et al* (including at 300 K). In the prevailing resonant conditions of the *un*pumped circumstance, the metastable $^{10}Cu_{III}^{2-}$ negative-$U$ state population, to which the effect has been attributed in [12–§2], can be expected to be appreciably smaller, and it is from this base that the $^{10}Cu_{III}^{2-}$ population then temporarily becomes built up further during relaxation from the laser pumping. As noted at the outset, the large size of the $^{10}Cu_{III}^{2-}$ entity comes from its internal charge imbalance, assisted by the fact that all the pd$\sigma$* antibonding states now have become occupied. In [1] Gedik *et al* introduce a lattice energy calculation based on their interpretive notion of a p-to-d 'charge transfer' excitation to access $(p^5)^{10}Cu_{II}^{1-}$ (often seen written briefly as $Ld^{10}$). They demonstrate that the rate at which the lattice parameter is observed to swell with the photon fluence delivered by a given laser pulse is precisely as follows from employing simple electrostatics. However, note that this same agreement would extend to our current negative-$U$ modelling, given the appearance of just half as many excited centres but these now bearing double the electronic charge. It should be restated here that Gedik *et al* concentrate upon the *c*-axis changes they are inducing, whilst Röhler, so as to avoid potential complications with Jahn-Teller effects and other structural detail, focussed upon the *ab* basal areas. In consequence of the *c*-axis softness of the present layered structures, some fraction of the above difference present between basal and *c*-axis percentage changes inevitably is going to reflect this distinction.

**§4. Towards a full understanding of the data from Gedik *et al.***

It is well at this juncture to be aware of an important distinction too between Gedik *et al*'s current pump/probe experiment and the earlier spectroscopic pump/probe work from Stevens *et al* [2]. There the electrons were pumped by the $Ti^{3+}$:sapphire laser not using the fundamental emission line, but its second harmonic, of photon energy 3.1 eV. After such excitation the $^{10}Cu_{III}^{2-}$ metastable population builds appreciably more rapidly under phonon- and electron-assisted trickle-down from the higher energy. With the Gedik arrangement, where the fundamental is employed, there is, by contrast, barely sufficient energy to reach the energy bar set by $^{9}Cu_{III}^{1-}$ for any ready advancement over into the doubly-loaded negative-*U* condition. Gedik *et al* accordingly experience a sizeable 'rise time' on the order of 100 ps for the metastable expansion to plateau out, prior to its subsequent and still slower nanosecond decline.

While this 'rise time' in the appearance of the pumped metastable condition is readily understandable, the reason for the observed fluence threshold regarding the structural modification demands more consideration. First remember what the above experiment records is not the population change *per se*, but lattice parameter change in response to that population change. If some new metastable population can be accommodated within precisely the same structural framework as a base population, whether natural or laser induced, there is no lattice parameter change to record. The framework within which the $^{10}Cu_{III}^{2-}$ materialize in post-pumped $La_2CuO_{4+\delta}$ is anticipated to be much like the 'stripe structure' existing in LBCO and LSCO. If the present author is correct in regard to the latter, the valence ordering structure acquired there actually takes 2-**q** diagonal form [12] and not the 1-**q** geometry invariably portrayed. The fact that the recent spin-polarized neutron diffraction data from Christensen et al [24] will indeed support the 2-**q** (spin) structuring advocated in [12] has very recently been attested to by Fine [25]. The 2-**q** charge stripe structure in [12] is of alternating $^{8}Cu_{III}^{0}$ and $^{9}Cu_{II}^{0}$ sites defining strings – the 'rivers of charge' – these bounding diagonal square domains of the less metallic, more magnetic, majority species, $^{9}Cu_{II}^{0}$. Within this structural order the most likely creation and dwelling spot for the $^{10}Cu_{III}^{2-}$ double-loading, whether spontaneous (persuant upon the above elaborated resonance) or laser induced, falls at the stripe crossing-points where the Madelung potential becomes greatest. Stripe formation is observed to be best organized at the sub-optimal hole content of $p = 1/8$, and at that stage there arise two stripe crossing-points within each $8a_o$ supercell [see 12–fig. 1]. By the $p = 0.155$ of optimal doping (in respect of $T_c$, although not the condensation energy) both these sites will take on the additional holes (*i.e.* have become $^{8}Cu_{III}$) [see 12–fig. 4], provided, that is, one stays with the $8a_o$ supercell – which is what the diffraction results would indicate for this composition.

Now of the threshold photon fluence (when each of the 64 copper centres within this cell has been delivered one eighth of a photon) only that fraction taken up around the $^{8}Cu_{III}^{0}$ centres in the stripes is able to engender the appearance of an excited $^{9}Cu_{III}^{1-}$. This could well mean then the production of only one such entity per cell. Only for fluences greater than this may the disproportionation/collisional process of equation (2) be able to secure in additive fashion the appearance of new doubly-loaded $^{10}Cu_{III}^{2-}$. The added fluence now witnesses this laser-induced population advance steadily. What such build up in $^{10}Cu_{III}^{2-}$ population (beyond that naturally present through process (4)) will mean for the superconducting properties of the system is an important one now to examine. It is not a foregone conclusion as to whether the additional local pairs induced by laser pumping are going actually to increase or decrease $T_c$. Recall, in the present resonant circumstances of the negative-$U$ modelling considered, there is delicate interplay between the local pairs and the sustained Cooper pairs. It is presumed that the absence of structural change prior to the emergence of the extra $^{10}Cu_{III}^{2-}$ content comes as a result of there being negligible alteration to the lattice conditions across process (1), the means wherewith the feed stock of $^9Cu_{III}^{1-}$ is procured: remember both $^9Cu_{II}^{0}$ and $^9Cu_{III}^{1-}$ are Jahn-Teller active.

The likely outcome is that the extra $^{10}Cu_{III}^{2-}$ emerging under process (2) will become deleterious to the superconducting performance if and when they should become heavily relaxed onto the lattice, and to impede thereby the free exchange of paired electrons with the sustained electronic environs. This is to say not only do the binding energies of the negative-$U$ pairs and the propagated Cooper pairs have to be quite similar, but so too do the geometric lattice environments in which the two interchanging varieties of pair find themselves based. If the local pairs are given the opportunity to relax too strongly onto the lattice, let alone segregate, this will impair not only the superconductivity but the basic conductivity of the mixed-valent system as well. It is the comparable problem in regard to heavy polaronic and then bipolaronic behaviour [26] which weighs heavily against a purely lattice-mediated sourcing of HTSC, vis-à-vis the present dominantly electronic scenario.

**§5. The support offered by further recent experiments to the above interpretation.**

A clear expression of what is happening in the Gedik experiment is afforded by the pump/probe spectroscopic data released a little while ago by Kusar *et al* [3] on LSCO, a study in which considerably lower laser light fluences were used. Working from the superconducting state, the latter group monitored the rise and subsequent relaxation of the excitation products by means of the changes incurred to the reflectivity at 1.5 eV. As a result of the very different rise and recovery behaviour of the component signals as a function of time, temperature and doping level, it proved possible to distinguish three very different channels to the relaxation. Primarily one senses recovery in the Cooper pair population, characterized by a relatively slow bottle-neck, phonon-controlled disruption and restitution – all of which action ends of course at $T_c$. Then there are in evidence the much more rapid changes to a further very substantial portion of the recovery signal. These one is able to link with the local pairs by their association with a $T^*$ that lies well in excess of $T_c$, and a $\Delta^*(T)$ that likewise is well in excess of $\Delta_c(T)$. Interestingly both latter gaps emerge from the analysis as relating specifically to the maximal d-wave values along the saddle, gap-antinodal, directions. Besides these two preceding components there has long been recognized to arise a third relaxation product [2,3], displaying an even longer relaxation period of nano- if not microsecond duration. It is this third component which now the much higher irradiation fluences employed by Gedik *et al* would seem to be promoting. The emergence of a great many $^{10}Cu_{III}^{2-}$ entities, far in excess of their natural equilibrium level, would appear to have led to enhanced self-trapping of these products. It is the local structural expansion of such temporary accumulations of $^{10}Cu_{III}^{2-}$ that one would look to be following in the Gedik experiment.

The above form of interpretation would stand in line with what has been reported at equilibrium within samples of steadily decreasing doping. Under their less metallicized, more ionic circumstances, as the $Cu_{III}$ centres take on greater local definition, the effect is to adjust the negative-$U$ level to somewhat greater binding energy $U$. This will diminish the electron-pair transfer rate between the $Cu_{II}$ and $Cu_{III}$ subsystems: as $\Delta^*$ rises, $\Delta_c$ falls [14]. The scanning tunnelling microscopy images very recently released by Gomez and colleagues [27] once more have very graphically illustrated the inhomogeneous nature to the instigating pairing. Those inhomogeneous environments, in BSCCO largely statistical in form, include both optimally disposed and pair-localizing $Cu_{III}$ regions. The latter localities grow in degree and extent with underdoping – and now, it would appear too, with strong laser pumping as well. The production of such expanded nanoregions is found in the pump/probe work to be virtually temperature invariant, but their temporary effect upon $T_c$ could well be very substantial, a matter that one now should try to determine.

With regard to matters dealing with pump/probe work and the negative-$U$ interpretation of HTSC, attention finally is directed to a very recent paper from Kaindl *et al* [28]. There the spectroscopic probe employed lies not in the eV but the meV range; *i.e.* in the 1-2 THz range. By ascertaining the changes to $\sigma_1(\omega)$ and $\sigma_2(\omega)$ at such energies issuing from the 1.5 eV pumping, Kaindl *et al* demonstrate, with the aid of a two-fluid model, in conjunction with the Rothwarf-Taylor rate equation for the relaxation, that the latter actually is dominated within the main 3-30 ps range by the term there *quadratic* in n*, the photo-excited, freed quasi-particle density. This observation then would endorse the route to the rapid re-establishment of pairing as indeed proceeding in bimolecular fashion as expressed above by equations (2) and (4), and which of course entail the appearance of the electronic negative-$U$ state $^{10}Cu_{III}^{2-}$.

We have reached accordingly a very feasible model with which to comprehend each point being presented by Gedik and colleagues in their most valuable of additions to HTSC phenomenology. It will be extremely interesting now to see whether those advocating spin fluctuation and phonon scenarios for HTSC find themselves in a position to address these striking new results.

**Acknowledgements**
The author is grateful to the University for receipt of a Senior Fellowship to permit this work to be carried forward, and to the members of the Low Temperature group here for continued discussions and support in advancing these matters.

,

**Figure caption**

**Figure 1**. The resonant negative-$U$ model from [9]. The symbols indicate a local coordination unit's nominal cation valence along with its instantaneous charge loading, both current and as deviation from the norm. The figure indicates the fluctuational state-loading energies for the trivalent ('doped') site loading sequence $^8Cu_{III}^0$, $^9Cu_{III}^{1-}$, $^{10}Cu_{III}^{2-}$ to be a re-entrant one. This is due to the shell closure at $p^6d^{10}$ and accompanying collapse in $\sigma/\sigma^*$ p/d interactions. Because of some permanent, metallizing, charge transfer between the two subsystems in the mixed-valent ground state, $^9Cu_{II}^0$ and $^8Cu_{III}^0$ are effectively brought to a common Fermi energy. In optimally doped HTSC systems the negative-$U$ double-loading fluctuational condition $^{10}Cu_{III}^{2-}$ becomes stationed in near-resonance with $E_F$. This makes $U_{eff}$ (per electron) = -1.5eV, in keeping with the value extracted from the femtosecond laser pump-probe optical experiments [2,3].

Note the customary definition of $U$ is $(E_{n+2} - E_{n+1}) - (E_{n+1} - E_n) = [E_{n+2} + E_n] - 2E_{n+1}$ ,

which in the present case gives $U = [E(d^{10}) + E(d^8)] - 2E(d^9)$ ,

and where note $\quad U = U_{effective} = U_{Coulombic} - U_{shell\ closure}$ .

Given $U_{Coul}$ = +3eV, and with $U_{shell\ closure}$ = -6eV, we reach $U_{eff}$ = -3eV per pair or -1.5eV per electron, the value supported by the experiments of references [2-5].

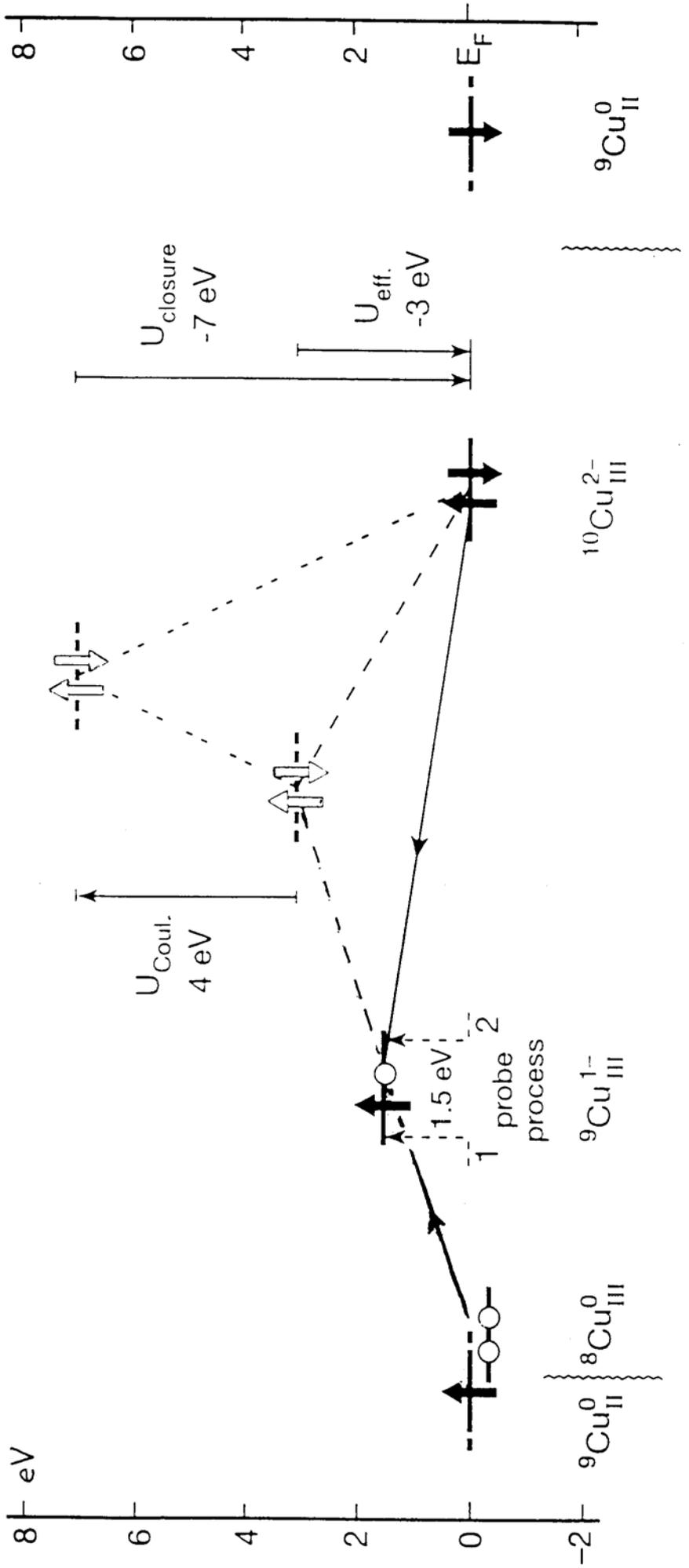

15